\documentclass{article}

\usepackage{PRIMEarxiv}

\usepackage[utf8]{inputenc} 
\usepackage[T1]{fontenc}    
\usepackage{hyperref}       
\usepackage{url}            
\usepackage{booktabs}       
\usepackage{amsfonts}       
\usepackage{nicefrac}       
\usepackage{microtype}      
\usepackage{lipsum}
\usepackage{fancyhdr}       
\usepackage{graphicx}       
 
\graphicspath{{media/}}     

\title{Building Ears for Robots: Machine Hearing in the Age of Autonomy
}

\author{
  Xuan Zhong \\
  Zoox, Inc. \\
  Foster City, California, USA\\
  \texttt{xzhong@zoox.com} \\
}

\begin{document}
\maketitle

\begin{abstract}
This study explores the significance of robot hearing systems, emphasizing their importance for robots operating in diverse and uncertain environments. It introduces the hardware design principles using robotaxis as an example, where exterior microphone arrays are employed to detect sound events such as sirens. The challenges, goals, and test methods are discussed, focusing on achieving a suitable signal-to-noise ratio (SNR). Additionally, it presents a preliminary software framework rooted in probabilistic robotics theory, advocating for the integration of robot hearing into the broader context of perception and decision-making. It discusses various models, including Bayes filters, partially observable Markov decision processes (POMDP), and multiagent systems, highlighting the multifaceted roles that robot hearing can play. In conclusion, as service robots continue to evolve, robot hearing research will expand, offering new perspectives and challenges for future development beyond simple sound event classification.

\end{abstract}


\section{Introduction}

Robot hearing system is an interesting topic of research in the fields of both hearing science and robotics. However, compared to computer vision, robot hearing systems received limited attention so far due to the scarcity of robots in uncertain and open environments. The field of robotics is still in its infancy and cannot deal with the complexity and uncertainty of the environment. As a result, most robots could only operate in highly controlled environments, such as automobile factories. However, in recent years, several types of service robots, such as robotaxis, have been deployed. The field of robotics is also quickly evolving with the advent of modern techniques like deep reinforcement learning. In the coming decades, as robots increasingly coexist with humans in open environments, the significance of robot hearing systems will grow, as they uniquely contribute to robot perception.

The current body of literature on robot hearing topic comes from several different fields. Significant progress was made in the field of sound event localization and detection (SELD), first as summarized in \cite{virtanen2018computational}, then in several years of DCASE competitions. The work has been focused on the application of machine learning methods in computational auditory scene analysis. The state-of-art audio perception models can detect, locate, and classify multiple sound sources in noisy three-dimensional (3D) environments \cite{kong2020panns}. While many SELD methods apply to robots, the meaning of sound only partially depends on the audio features. Robot hearing systems should encompass capabilities beyond simple sound event classification.

Another area of study that is related to robot hearing systems is the comparative study of human and machine hearing. Human auditory perception has been studied through physiology and behavioral psychology experiments, and mathematical models were established to mimic the behaviors of human auditory systems \cite{pastore2020cross}. Such models could be potentially applied to robots. The seminal work of \cite{lyon2017human} started with an insightful review of human hearing research and the theory of signals and systems. The focal point of this book is the CARFAC (cascade of asymmetric resonators with fast-acting compression), a computational model of audio features that generates a rich representation of input auditory signals for high level processing. The current article seeks to address unresolved questions in machine hearing within the field of robotics.

The current study starts with a discussion on robot hearing hardware systems. Section 2 delves into the hardware design of robot ears, using exterior microphone arrays on self-driving cars as an example. It covers system design principles, goals, challenges, methods, and testing procedures, with a focus on achieving a suitable signal-to-noise ratio (SNR) amid the robot's noise and environmental factors.

Section 3 delves into the roles of robot hearing systems within the modern probabilistic robotics framework, covering Bayes filters, partially observable Markov decision processes (POMDP), and multiagent systems. Despite the emerging nature of robotics, the study underscores the established interactions between agents and their environments. It suggests that integrating observation and action opens up new avenues for the design and research of robot hearing systems. Moreover, it underscores the limitations of current sound event localization and detection (SELD) approaches, emphasizing the importance of considering the temporal dimension of sound events in relation to sequential observations and decision-making. This temporal context, or the timing of sound events relative to agent actions, is deemed crucial for a more comprehensive sound analysis.

\section{Robot Hearing Hardware Design }
\label{sec:headings}

This section presents the goals, challenges, design guidelines and test methods of the hardware of robot hearing systems. Throughout each subsection, the discussions use the robotaxi exterior microphone arrays, one of the first mass-produced robot hearing systems, as examples.

\subsection{Goals }
Similar to human ears, the hardware of robot hearing systems captures sound waves and preprocesses the audio signals before sending them to the higher levels of the perception systems.

\begin{figure}[h]
\includegraphics[width=0.8\textwidth]{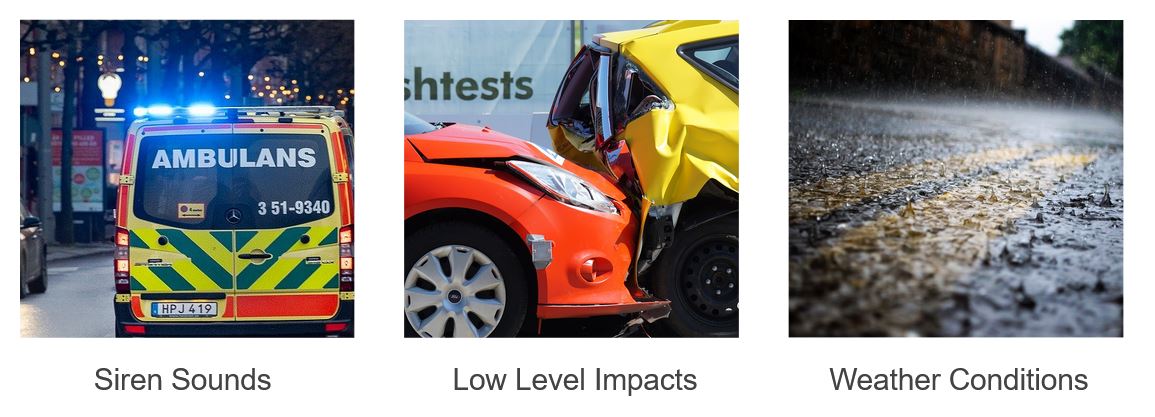}
\centering
\caption{Applications of machine hearing systems on autonomous vehicles}
\centering
\end{figure}

Exterior microphones on autonomous vehicles serve multiple important purposes (Figure 1). Firstly, they are essential for reliable siren detection and localization, ensuring that autonomous vehicles can respond appropriately to emergency vehicles, especially in blind corners where traditional sensors may not provide adequate information.

Secondly, in the event of a minor accident with low level impact, exterior microphones capture impact sounds. This is essential for reporting and assessing injuries or damages accurately, contributing to efficient post-accident procedures.

Furthermore, autonomous vehicles also need to monitor the local weather conditions to evaluate the possibilities of sensor degradation. Rainfalls and many other weather conditions carry special audio features that can be used for weather event classification.

\subsection{Challenges}

Designing effective robotic hearing systems presents a significant challenge: converting sounds into audio signals with a sufficient SNR for high level processing such as sound event detection. This challenge is compounded by factors related to both the robot and the environment.

In autonomous driving scenarios, robotaxis generate internal noise sources, such as engine and air-conditioner sounds, which can obscure acoustic features, especially for low-level signals such as distant sirens. Additionally, the robot's physical size can create sound reflections and diffractions, leading to substantial sound attenuation in directions not directly facing incoming sound waves. To address this, multiple microphone arrays can be placed on different sides of the vehicle to enhance coverage. Furthermore, accommodating the robotic hearing system alongside other modules and subsystems within the vehicle necessitates thoughtful packaging and harnessing considerations.

Environmental factors further complicate the hardware design of robot hearing systems. For exterior microphones on robotaxis, wind noise increases with higher vehicle speeds, and traffic noise can diminish the SNR. These microphones are also exposed to environmental elements like water and dust, making it challenging to strike a balance between wind noise reduction and protecting against ingression. Moreover, for sound event detection at long distances, sounds may not propagate along a straight line due to local atmospheric conditions like temperature gradients and local winds \cite{hannah2006ground}.

In summary, calculating the SNR of exterior microphones on robots must account for the variability of both the signal and the noise. The noise level is determined by the most dominant noise sources within the environment and originating from the robot itself. 

\subsection{Design Methods}

Given the complexity of both the robot itself and its operating environment, the hardware design of robot hearing systems should follow a system design approach \cite{hamming1998art}. According to Hamming, system design involves keeping the high level objectives in mind and translating local design efforts into global outcomes. Several key principles can guide this approach:

\begin{enumerate}
\item \textbf{Understand the Whole as a Whole}: Comprehend the system as a holistic entity, not merely the sum of its individual components.

\item \textbf{Flexibility and Adaptability}: Prepare for potential changes and adapt to evolving circumstances.

\item  \textbf{Subsystem Specifications with Redundancy}: Define subsystem specifications with built-in redundancies to enhance reliability.
\end{enumerate}

In the case of designing robotaxi exterior microphones, using a system design approach requires the following:

\begin{enumerate}
\item \textbf{Top-down Design Approach}: Define the vehicle system level performance before specifying subsystem and module performance.

\item \textbf{General Purpose Design}: Design a general purpose microphone for a wide range of tasks instead of, for instance, building a resonator that only picks up a narrow range of target sounds.

\item \textbf{Hierarchical Design with Redundancy}: break the top level design requirements to low level design targets while keeping some redundancies.
\end{enumerate}

\begin{figure}[h]
\includegraphics[width=0.8\textwidth]{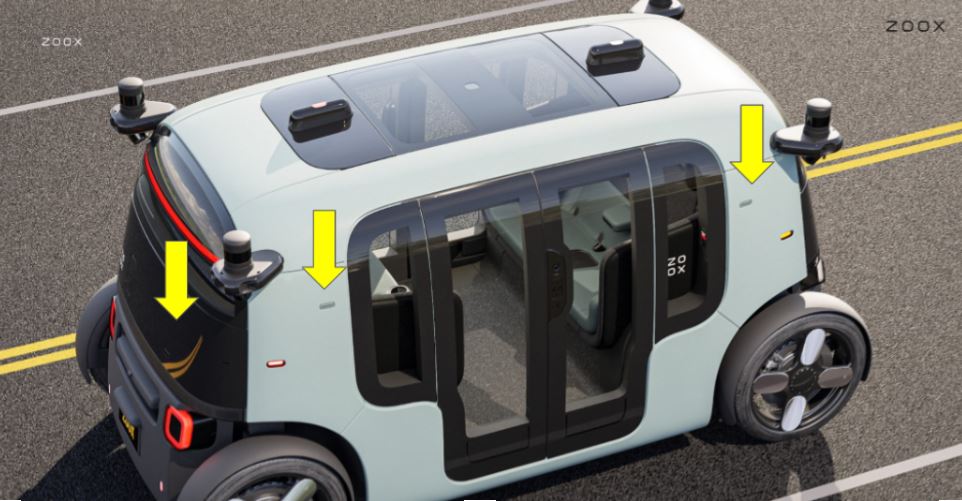}
\centering
\caption{Placement of multiple exterior microphones on a Zoox autonomous vehicle}
\centering
\end{figure}

The Zoox robotaxi's exterior microphone system embodies the design principles discussed earlier. It comprises six pairs of microphones strategically positioned around the vehicle's perimeter (Figure 2). This placement is informed by extensive field and lab tests focused on vehicular and wind noise. At the microphone module level, the detailed design is inspired by the feather structure of barn owl \cite{koch2002morphometry}, which finds a balance among acoustics, wind noise reduction and ingression protection performance. The design delivers wide-band audio signals at high speeds. The design passes the IP69k test, which means that the microphones are fully dust-proof and can withstand water jet streams of 1500 PSI at a temperature of 80 °C.

\subsection{Test Methods}

Two types of tests are critical to the success of product development. For the design team, a series of validation tests are required to ensure that the design could fulfill the required tasks. At the factory, a battery of quality control tests is implemented, starting from the component level and progressing upward to ensure consistent adherence to specifications.

For example, frequency responses of microphones are tested at various levels. The MEMS microphone components are tested before they are soldered onto the printed circuit board assembly (PCBA). Further testing occurs at the PCBA level. Once the PCBA is integrated into enclosures, module-level testing is performed, with each test requiring specialized fixtures.

After the module level tests, the fully assembled modules are mounted onto a limited number of vehicles for validation tests. The frequency response is tested in two settings. In one test, the reference microphone is mounted on the vehicle, right next to the device-under-test. This “local” frequency response represents the capability of the microphone to capture nearby sounds, such as human speech or low-level impact. In another test, the reference microphone is used to calibrate a test loudspeaker without the vehicle. This "global" frequency response represents how good the microphone is at capturing distant sounds, such as the sirens. The results from multiple loudspeaker locations with different incident angles can also be named vehicle-related transfer functions, akin to the head-related transfer functions for human listeners.

\begin{figure}[h]
\includegraphics[width=0.8\textwidth]{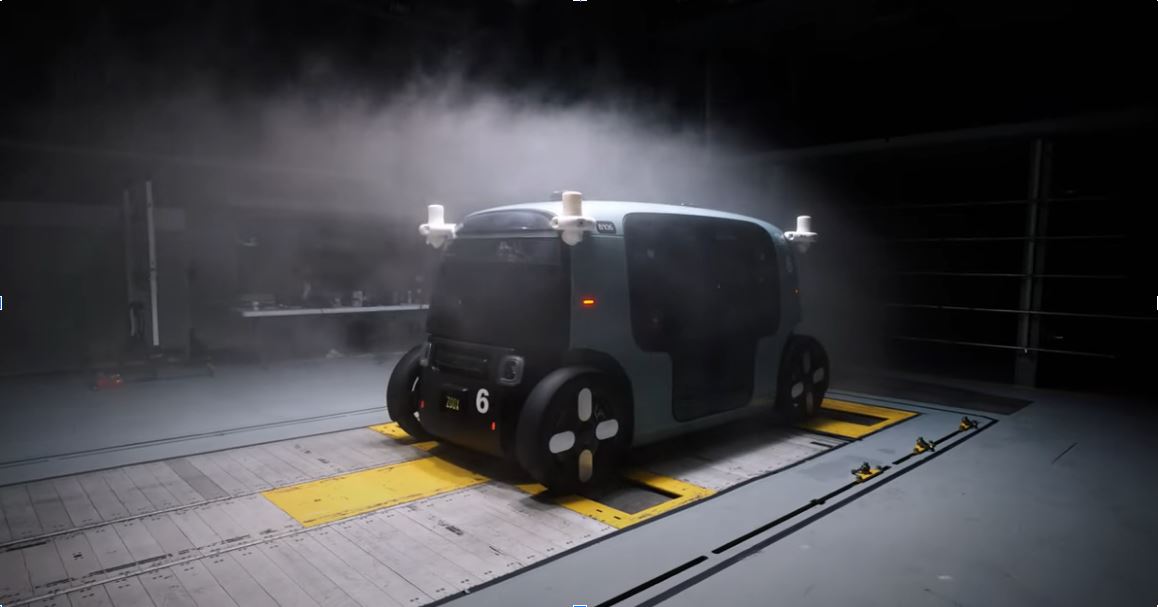}
\centering
\caption{Simulated rain test in a semi-anechoic wind tunnel}
\centering
\end{figure}

The wind noise tests also follow the “bottom-up” sequence. Airflow is generated at various scales to provide test environments at different levels. A small wind box with a 30-cm test section is used to study the wind noise reduction effect of different surface conditions. A medium-sized wind tunnel with a 5-m test section offers the capability of wind-blown rain testing at the module and sub-assembly levels. At the vehicle level, the lab tests leverage a big wind tunnel with a 20-m test section and weather simulation capabilities (Figure 3). Moreover, vehicle-level field tests are done at multiple test tracks to provide data of microphone performance with real winds.

\section{Robot Hearing Software Framework}

This section presents a preliminary software framework of robot hearing systems based on probabilistic robotics theory. This is an attempt to answer the question “what does a hearing system mean to a robot?”  rather than the specifics of system implementation.

\subsection{Space and Hearing}

Mostly of the previous literature on human spatial hearing has emphasized "hearing" rather than "space". Typically, a stationary listener listens to stationary sound sources, so that spatial hearing cues, such as interaural time difference (ITD), interaural level difference (ILD) and human-related transfer function (HRTF), could be studied \cite{blauert1997spatial}. The result is a comprehensive but at best partial understanding of the relationship between space and hearing. For instance, a recent study pointed out the limitations of head-centric spatial hearing cues in world-centric sound source localization tasks \cite{yost2021sound}.

In the field of robotics, intelligent agents, including humans and robots, navigate their environments with at least six degrees of freedom, sometimes sharing space with other agents. For example, a human listener is searching for a lost phone at home. If the lost phone is set to vibration mode, low-frequency sound sources are challenging to locate accurately. Hence it is unlikely that the listener can locate the phone at once. However, by moving around, the listener perceives variations in sound intensity. Continuously moving to maximize sound intensity eventually leads the listener to discover the phone on the couch.

\begin{figure}[h]
\includegraphics[width=0.2\textwidth]{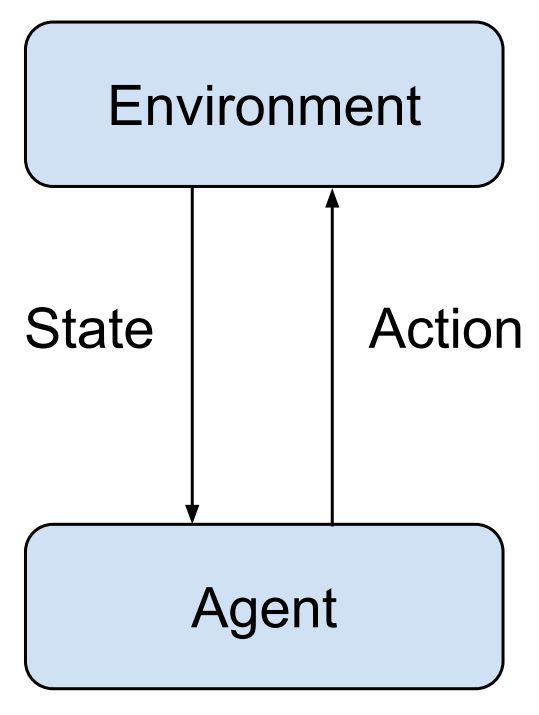}
\centering
\caption{The interactions between an intelligent agent and its environment}
\centering
\end{figure}

This example illustrates the use of a hearing system in a setting of probabilistic robotics \cite{burgard2005probabilistic}, in which the problem of finding a sound source is framed as a long-term state-estimation problem based on sequential observations and actions (Figure 4). In this context, the concept of "space" discussed earlier is referred to as the "environment". 

Theoretically, a state in probabilistic robotics encompasses all information pertaining to the future in the environment and the agent itself. In practice, a selective subset of these variables is calculated to optimize time and computational resources.

Intelligent agents engage with their environments in two ways: through actions performed by actuators and through observations made using sensors, including microphones. It's important to note that this study doesn't assume that an agent can accomplish its tasks solely with audio inputs.

Many existing robot perception systems use an open-loop design, in which the role of the system is to generate a percept regarding nearby objects and events without considering how these perceptions inform decision-making. The majority of the current SELD research and development falls in this category and they can fulfill a wide range of tasks. 

However, there are distinct advantages to integrating perception with planning within the framework of probabilistic robotics. Hence the remainder of this section focuses on the discussion of the closed-loop design of sequential decision-making processes. The taxonomy of robot perception and decision-making is introduced in the order of Bayesian filters, partially-observable Markov decision processes (POMDP) and multiagent systems, with increasing complexity and capabilities. The exploration of the potential roles of robot hearing within these systems follows.

\subsection{Robot hearing in Bayes filters}

The agent interacts with its environment through a sequence of observations and actions that changes the state. The sequence of these state changes is illustrated in Figure 5.

\begin{figure}[h]
\includegraphics[width=0.8\textwidth]{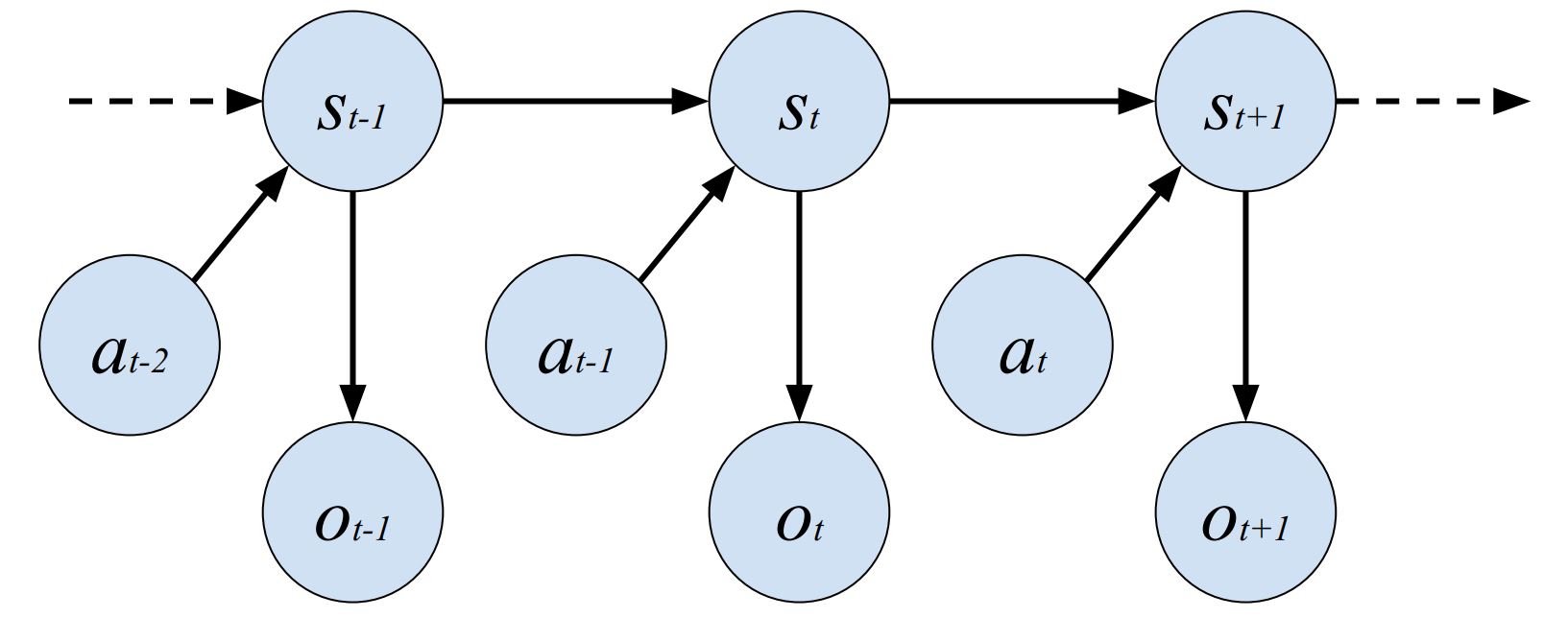}
\centering
\caption{The probabilistic model of state changes over time}
\centering
\end{figure}

In probabilistic robotics, the state is described by a probabilistic density function rather than a deterministic function. The sequence of interactions consists of measurement updates and action updates. Action updates typically add to the uncertainty due to the error of the actuator and environment uncertainties (similar to walking in darkness), whereas the measurement update usually reduces the uncertainty by providing added information \cite{burgard2005probabilistic}.

The agent does not have direct access to the true state but must infer the state through a series of actions and observations. Observations can come from a range of sensors, including cameras, lidars, radars and microphones. The agent’s internal knowledge of the state is referred to as belief. The general approach of deriving belief from actions and observations is the Bayes method, in which the belief converges to the true state after a series of actions and observations with sensors and actuators with Gaussian noise. The errors reduce significantly as more observations and actions are incorporated.

The transition model describes how the dynamics of the environment changes based on the previous state and action: \(s_t \sim p(s_t|s_{t-1}, a_{t-1})\)

The measurement model describes the partial sensor observation of the state: \(o_t \sim p(o_t|s_t)\)

The goal is to estimate the state from sensor data and controls.

The most studied form of Bayes filters is the Kalman filter and its nonlinear version, the extended Kalman filter (EKF). In a previous study \cite{zhong2016active}, robot head rotation and interaural time differences (ITD) are combined in an EKF model for sound source localization. Here the state represents the location of the sound sources, the observations consist of the binaural audio inputs and ITDs, and the actions correspond to the head rotations.

When sequential observations are allowed during head rotations, the amount of ITD change becomes a reliable new cue for sound source localization, allowing the robot to estimate not only the azimuth of the sound source, but the elevation as well, although at any given moment, the ITD alone does not support sound source localization due to the problem of the cone of confusion. 

As another example, in the case of “finding the phone”, the listener can potentially locate the phone even with only one ear. The monaural sound level changes over a trajectory of motion are another spatial hearing cue. In fact, for the simplest model of binaural hearing, which is two microphones attached to the two ends of a rod, all the degrees of freedom contribute to spatial hearing except the rotation along the axis of the rod.

\subsection{Robot hearing in POMDP}

In the Bayes filter model, a control of action is given for each step in the sequence of state changes. However, the Bayes model itself does not prescribe how the control is generated. In the realm of probabilistic robotics, the basic framework of planning and control is the Markov decision process (MDP). A more general form of MDP is the POMDP (Figure 6), where the sensors only carry partial information about the states \cite{burgard2005probabilistic}.

The primary objective of robot software is to select the right actions. However, when a robot interacts with the environment, there are two types of uncertainty that a robot model must consider. The first type of uncertainty pertains to the sensors. At best, each type of sensor only generates signals that reflect a part of the true state with some noise. The second type of uncertainty concerns the actuators. Due to the actuator noise and the complexity of real-world dynamics, the effects of actions are not deterministic but probabilistic in nature.

\begin{figure}[h]
\includegraphics[width=0.8\textwidth]{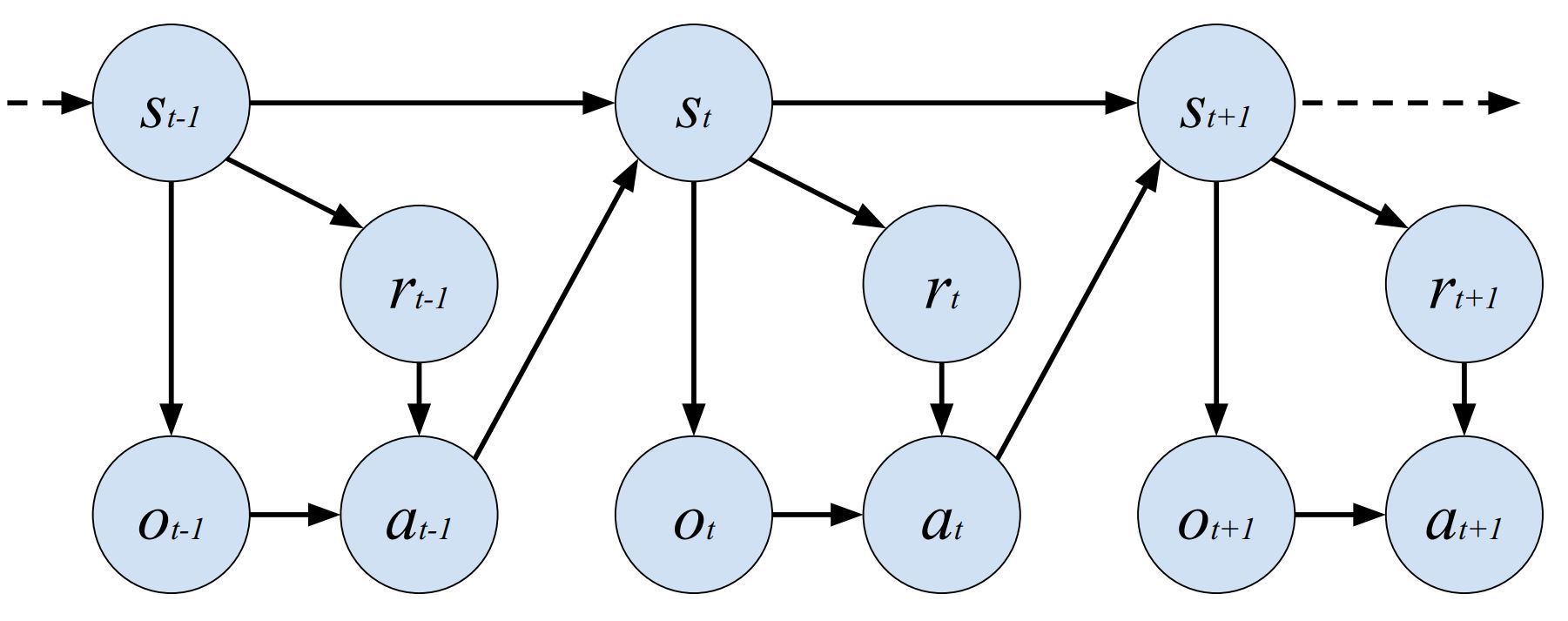}
\centering
\caption{Partially-observable Markov Decision Process (POMDP)}
\centering
\end{figure}

The POMDP model formulates the problem as the following:

\-\hspace{12mm} \(s\): set of states 
    
\-\hspace{12mm} \(o\): set of observations
    
\-\hspace{12mm} \(a\): set of actions
    
\-\hspace{12mm} \(r\): rewards

The decision process is modeled as follows:

\-\hspace{12mm} Transition model: \quad \quad \(s_t \sim p(s_t|s_{t-1}, a_{t-1})\)

\-\hspace{12mm} Measurement model: 	 \quad   \(o_t \sim p(o_t|s_t)\)

\-\hspace{12mm} Reward function:  \quad \quad  \(r_t \sim p(r_t|s_t)\)

\-\hspace{12mm} Policy: \quad \quad \quad \quad \quad \(a_t \sim p(a_t|o_{t-1}, a_{t-1})\)

The primary aim of the POMDP model is to determine a policy of actions that maximizes the cumulative reward in the process. Solving POMDPs can be challenging, and various methods are employed, including traditional techniques such as value iterations \cite{burgard2005probabilistic} and modern approaches like deep reinforcement learning networks \cite{wu2023daydreamer}. A comprehensive overview of reinforcement learning methods is available in \cite{dong2020deep}. 

A properly solved POMDP model automatically regulates the behavior of observations and actions. The decisions of the policy at a given time step can be taking an action or “doing nothing and observing”. When the uncertainty is high, the agent keeps observing and collecting more information. On the contrary, when the uncertainty is low, the agent takes actions according to the policy. There could also be a process of exploration, in which the only goal is to collect more information about the environment through active sensing \cite{burgard2005probabilistic}.

The roles of robot hearing system within the POMDP framework is multi-faceted:

\begin{enumerate}
\item \textbf{Providing Information about the States:} Sensors should provide information to generate a proper belief state for POMDP. Each type of sensor reflects a different aspect of the state. Hence robot hearing reduces belief variance by reflecting some information of the state. Hearing does not require significant computing power and is a rapid mode of perception. Hearing also does not require direct field of the view of the sound sources, so it is beneficial for state estimation when the field of view is obstructed for other sensors such as cameras. In many cases, what blocks the field of view is the enclosure of objects under observation. Sounds of diesel engines and hair blowers provide information about the states which may be difficult to obtain with other methods.

\item \textbf{Assessing Effects of Actions:} Robot hearing also plays a role in evaluating the consequences of previous actions. For example, when a car is stared, the sound of the engine can indicate whether it started normally or remained silent due to a dead battery. Similarly, when a door is closed, the sound of the lock can reveal whether it engaged as expected. In the context of POMDP, silence can hold meaning: it signifies an anticipated state change that did not occur after the actions were taken. Consequently, the meaning of sound events does not only depend on the audio inputs, but also the context of interactions between the robot and its environment in the decision process.
\end{enumerate}

\subsection{Multiagent systems}

The concept of multiagent systems introduces several complexities and challenges into the realm of robotics and artificial intelligence. In multiagent systems, two or more agents coexist and interact within the same environment, each striving to maximize their individual rewards or objectives. This introduces a level of complexity beyond what is typically addressed in traditional POMDP settings, where a single agent operates in a more controlled environment. \cite{shoham2008multiagent}.

Multiagent systems can be fully cooperative, fully competitive, or a mix of both. They can be modeled as Markov games, where all agents decide simultaneously, or extensive-form games, with turn-based decisions. Agents have limited information about the state and lack access to other agents' policies, making it challenging to predict their actions and effects.

In the realm of multiagent systems, robot hearing systems fulfill several critical roles in addition to those in the case of POMDP. Firstly, they serve as detectors for identifying the presence of other agents in the environment. In dynamic scenarios, the number of agents can fluctuate as they enter or exit the immediate area of interest. Consequently, agents equipped with robot hearing have an enhanced capacity to detect the presence of new agents. This capability is essential for tasks such as recognizing potential mates, predators, or prey.

Additionally, robot hearing systems can monitor the actions of other agents. When other agents take actions, the environment may undergo changes that are independent of the primary agent's actions. In contrast, in moments of silence, the likelihood of external actions occurring is low, and the primary agent can reasonably assume that its existing policies remain applicable. The primary agent may also gain insights into the intentions, rewards, and policies of other agents by analyzing a series of their actions.

Robot hearing systems must take into account the broader context of multiagent decision-making to accurately interpret sound events. For instance, hearing the front door open may have very different implications depending on whether one expects a roommate or is home alone at night. Neglecting the influence of multiagent interactions on sound event interpretation can limit the applicability of existing methods in complex scenarios. This underscores the idea that sound events should be studied in conjunction with the context in which they occur, as their meaning is only partially determined by the audio inputs.

\section{Discussions}

This study first introduces the hardware design of robot hearing systems with the example of robotaxi. The design challenges, principles and test methods were discussed. Looking ahead, service robots may take a myriad of sizes and forms. There may be large robots for mining and very small drones for pollination, but the design and validation principles of the hearing systems should still apply.

The following section delved into the significance of a hearing system for a robot. Drawing from probabilistic robotics models, a hearing system contributes to improved belief certainty through multiple observations (in EKF), provides accurate and timely information about the state and action effects (in POMDP), and tracks environmental changes caused by other agents (in multiagent systems).

Indeed, this preliminary framework emphasizes the importance of not isolating the study of hearing systems but integrating them with actions and sequential decision-making within the broader field of robotics. This approach opens up several future research directions that extend beyond the current paradigms in both SELD and hearing research. In the future, as the functions, capabilities and complexity of service robots quickly evolves, so will their hearing systems.

\section*{Acknowledgments}

The author would like to thank Shaminda Subasingha, William Yost,  Aurora Cramer and Liang Sun for their comments and suggestions. The materials covered in this article were presented and discussed at the Hearing Seminar at Stanford University, which was organized by Malcolm Slaney in October, 2023.

\bibliographystyle{apalike}
\bibliography{referencesA}

\end{document}